\documentclass[12pt]{article}

\usepackage[hmargin=2.5cm,vmargin=3cm]{geometry}
\usepackage{graphicx}
\usepackage{amsmath}
\usepackage{wasysym}
\usepackage{enumerate}
\usepackage{graphicx}  
\usepackage{dcolumn}   
\usepackage{bm}        
\usepackage{amssymb}   
\usepackage{enumerate}
\usepackage{amsmath}
\usepackage{float}
\usepackage{multibib}
\usepackage{cite}
\begin{document}

\title{Classification theorem and properties of singular solutions to the Tolman-Oppenheimer-Volkoff equation}
\author {Charis Anastopoulos\footnote{anastop@physics.upatras.gr}  and  Ntina Savvidou\footnote{ksavvidou@upatras.gr}     \\
 {\small Department of Physics, University of Patras, 26500 Greece} }
\maketitle
\begin{abstract} The Tolman-Oppenheimer-Volkoff (TOV) equation admits singular solutions in addition to regular ones. Here, we prove the following  theorem. For any equation of state that (i) is obtained from an entropy function, (ii) has positive pressure and (iii) satisfies the dominant energy condition, the TOV equation can be integrated from a boundary inwards to the center. Hence, the thermodynamic consistency of the EoS precludes pathological solutions in which the integration terminates at finite radius (because of  horizons, or divergences / zeroes of energy density).
At the center, the mass function either vanishes (regular solutions) or it is negative (singular solutions). For singular solutions, the metric at the center is locally isomorphic to negative-mass Schwarzschild spacetime.
This means that matter is stabilized because the singularity is strongly repulsive. We show that singular solutions are causally well behaved: they are bounded-acceleration complete, and they are conformal to a globally hyperbolic spacetime with boundary. Finally,  we show how to modify unphysical equations of state in order to obtain non-pathological solutions, and we undertake a preliminary investigation of dynamical stability for singular solutions.

\end{abstract}

\section{Introduction}
The Tolman-Oppenheimer-Volkoff (TOV) equation describes a static, spherically symmetric matter configuration with gravitational self interaction. Regular solutions to the TOV equation provide the simplest models of compact stars (white dwarves, neutron stars), in which rotation can be ignored. They have been studied ever since the 1930s and their properties are well understood.

However, regular solutions to the TOV equations form a set of measure zero in the set of all solutions. Most solutions to the TOV equation are singular. With few exceptions \cite{ZuPa84, CoKa94, AnSav12, AnSav16, Kim17}, singular solutions  have been ignored in the bibliography, possibly because they are characterized by a naked singularity at the center. In this paper, we study these solutions with general equations of state (EoS) for matter. In particular, we analyze the TOV equation as an initial value problem, where the equation is integrated from an outside boundary (e.g., a star's surface)  inwards. In contrast, regular solutions are usually specified by conditions  both at the center (regularity) and at the boundary.


In this paper, we  focus on the analytic and geometric properties of singular solutions. We make no effort to argue about their physical relevance. We want to separate between the  mathematical facts about these solutions  and   their physical interpretation. The latter will be presented in a different publication.

The motivation for this work is  three-fold. First, we want to provide a classification of all singular solutions to the TOV equation. We find it quite surprising that such a classification is absent in the bibliography, given the fact that the TOV equation is a key paradigm of relativistic astrophysics. We show that all singular solutions share a common structure, including the geometry of the singularities.

Second, in Ref. \cite{AnSav12} we proposed that singular solutions to the TOV equation are essential for the thermodynamic consistency of  gravitating matter, even if they are viewed solely as virtual solutions. Thermodynamic consistency requires a consistent assignment of entropy to   the singularities of these solutions. This entropy assignment provides a concrete implementation of
 Penrose's conjecture about a relation between spacetime singularities and entropy \cite{Penrose1, Penrose2}. We believe that the result of Ref. \cite{AnSav12} can be generalized to timelike singularities in generic static spacetimes. The present classification of static spherically symmetric spacetimes is the first step towards such a generalization.

Third,   the solutions studied here may be important for understanding gravitational collapse. Spherically symmetric gravitational collapse leads to the formation  of naked singularities for generic  (spherically symmetric)  initial conditions \cite{ES79, Ch84}. The detailed properties of the naked singularities are model-dependent, and there is a long-standing discussion about their physical relevance---see,  \cite{Joshi1, Joshi2} and references therein. The   naked singularities considered in this paper do not involve non-extendible geodesics, and in
this sense they are much milder than the ones appearing in existing models. For this reason, it would be important to understand whether the solutions studied here can be obtained as end states of gravitational collapse.

A  reason that complicates the study of singular solutions to the TOV equation  is that the integration of the latter from a boundary inwards often terminates at  finite radius. There, the pressure diverges, or it vanishes, or a horizon is present. Such pathologies are common in many widely used EoS for matter, like the polytropic ones.

In this work, we show that these pathological behaviors are artefacts of thermodynamically inconsistent EoS.    A consistent EoS for matter must be derived from an entropy density function, subject to the fundamental thermodynamic axioms \cite{Call}. Many popular EoS employed in the study of compact stars are not thermodynamically consistent in this sense. They are designed in order to reflect a relation between pressure and density that is valid in a particular range of densities / temperatures. Outside this range, their behavior may be physically problematic.  We show that the pathologies that appear in the integration of the TOV equation originate solely from inconsistencies in the EoS.

We prove the following theorem. For any thermodynamically consistent EoS that satisfies the dominant energy condition $( P \leq \rho$) the TOV equation is always integrated up to the center. Hence, there are only two types of solution: regular at the center and singular at the center.

We find that all singular solutions share the same structure and they are characterized by a curvature singularity at the center. The latter is locally isomorphic to the singularity of the negative-mass Schwarzschild spacetime. The singularity repulses all matter in its vicinity, and this repulsion stops the collapse of the exterior layers.

We also analyse the causal structure of the singular solutions, and we find that if the singularity is treated as a boundary, then, the spacetime is causal-geodesic complete. In fact, it is conformal to a globally hyperbolic spacetime with boundary.

Finally, we  undertake a preliminary study of dynamical stability for singular solutions. We consider radial adiabatic perturbations. We find that the singularity, in general, enhances stability, and that instability is caused by surfaces of high blue-shift in the outer layers. We give reasons why we expect stable solutions to be generic, and we verify this expectation in a simple model.

The structure of this paper is the following. In Sec. 2, we describe the main background results on thermodynamics of gravitating systems. We also formulate  a precise integrability condition that must be satisfied by any thermodynamically consistent EoS. In Sec. 3, we derive our main result, the theorem that the TOV equation can be integrated up to the center, and we identify the common structure shared by all singular solutions. In Sec. 4, we analyse the causal properties of singular solutions. In Sec. 5, we consider the zero temperature limit of singular solutions, and we show how pathological EoS can be remedied by changing their low- and high-temperature behavior. In Sec. 6, we discuss stability under radial adiabatic perturbations. In Sec. 7, we discuss our results.

\section{Thermodynamics of gravitating systems}
\subsection{Key properties}
First, we summarize some thermodynamic properties of  gravitating matter in equilibrium \cite{KM75, SavAn14}.

 We  consider a static, globally hyperbolic spacetime $M = {\pmb R}\times \Sigma$ with four-metric
\begin{eqnarray}
ds^2 = -L^2(x) dt^2 + h_{ij}(x) dx^i dx^j, \label{4metric}
\end{eqnarray}
expressed in terms of the spatial coordinates $x^i$ and the time coordinate $t$. The time coordinate $t$ defines a spacelike foliation on $M$ in terms of spacelike surface $\Sigma_t$.
 $L$ is the lapse function, and $h_{ij}$ is a $t$-independent Riemannian three-metric on $\Sigma_t$. The time-like   unit normal on the foliation  is $n_{\mu}= L \partial_{\mu}t$ and the extrinsic curvature tensor on $\Sigma_t$ vanishes.

 Let  $C \subset \Sigma$ be a compact spatial region, with boundary $B = \partial C$. $C$ contains an isotropic fluid in thermal and dynamical equilibrium, described by the stress-energy tensor
\begin{eqnarray}
T_{\mu \nu} = \rho n_{\mu}n_{\nu} + P (g_{\mu \nu} + n_{\mu} n_{\nu}), \label{tumn}
\end{eqnarray}
where $\rho(x)$ and $P(x)$ are the energy density and the pressure, respectively.

 The continuity equation $\nabla_{\mu}T^{\mu \nu} = 0$ for the metric (\ref{4metric}) is
\begin{eqnarray}
\frac{\nabla_i P}{\rho +P} = - \frac{\nabla_iL}{L}. \label{cont}
\end{eqnarray}

We assume that the fluid consists of $k$  particle species. The associated particle-number densities $n_a(x)$, $a = 1, \ldots, k$, together with the energy density $\rho(x)$ define the thermodynamic state space. All local thermodynamic properties of the fluid are encoded in the entropy-density functional $s(\rho, n_a)$. The first law of thermodynamics takes the form
\begin{eqnarray}
T ds = d \rho - \sum_a \mu_a dn_a, \label{1st}
\end{eqnarray}
where $\mu_a = - T \frac{\partial s}{\partial n_a}$ is the chemical potential associated to particle species $a$ and $T = \left(\frac{\partial s}{\partial \rho}\right)_{n_a}^{-1}$ is the local temperature. The pressure $P$ is defined through the Euler equation
\begin{eqnarray}
\rho + P - Ts -\sum_a \mu_a n_a = 0. \label{euler}
\end{eqnarray}
 Combining  Eqs. (\ref{euler}) and  (\ref{1st}), we derive the Gibbs-Duhem relation,
$dP  = s dT + \sum_a n_a d\mu_a$.

The total entropy for matter is  given by $S = \int_C d^3x \sqrt{h} s(\rho, n_a)$, where $h$ is the determinant of the three-metric $h_{ij}$.
We maximize $S$ for fixed values of the total particle numbers in $C$, $N_a = \int_C d^3x \sqrt{h} n_a $. To this end, we vary the function
  \begin{eqnarray}
\Omega =  S + \sum_a b_a N_a
  \end{eqnarray}
 with respect to $n_a$, where  $b_a$ are Lagrange multipliers. $\Omega$ is a Massieu function obtained by the Legendre transform of entropy. We will refer to it as the {\em free entropy} of the system. The name is analogous to the free energies (Gibbs and Helmholtz) that are defined as Legendre transforms of the internal energy functions in thermodynamics.

 Variation with respect to $n_a$
\begin{eqnarray}
 \delta \Omega =   \sum_a \int_C d^3 x \sqrt{h} \left( -\frac{\mu_a}{T} + b_a\right) \delta n_a = 0,
\end{eqnarray}
leads to
 $b_a =  \frac{\mu_a}{T}$.  Hence, for equilibrium configurations the thermodynamic variables $\frac{\mu_a}{T}$ are constant in $C$. We will refer to   $b_a =  \frac{\mu_a}{T}$ as the {\em activity} of the particle species $a$. (The names "activity" and "fugacity" is sometimes employed for $e^{b_a}$.)

The free entropy density $\omega$  is the Legendre transform of the entropy density $s$ with respect to $n_a$
\begin{eqnarray}
\omega(\rho, b_a) := s -\sum_a \frac{\partial s}{\partial n_a} n_a= s + \sum_a b_a n_a = \frac{\rho + P}{T}. \label{omega}
\end{eqnarray}
Obviously,
\begin{eqnarray}
\Omega = \int_C d^3x \sqrt{h} \omega(\rho,b_a). \label{OM}
\end{eqnarray}

Substituting Eq. (\ref{omega}) into Eq. (\ref{1st}), we obtain
\begin{eqnarray}
 d \omega = \frac{d \rho}{T} + \sum_a n_a db_a.
\end{eqnarray}
It follows  that
$ T^{-1} = (\partial \omega/\partial \rho)_{b_a}$ and  $n_a =  (\partial \omega/\partial b_a)_{\rho}$. The Gibbs-Duhem relation becomes

\begin{eqnarray}
dP = \omega dT + T \sum_a n_a db_a.
\end{eqnarray}

For entropy-maximizing configurations, $db_a = 0$, hence,
 \begin{eqnarray}
\frac{ dP}{dT} = \omega = \frac{P+\rho}{T}. \label{dpt}
 \end{eqnarray}
  Combining with Eq. (\ref{cont}), we obtain
\begin{eqnarray}
\frac{\nabla_iT}{T} = - \frac{\nabla_i L}{L}, \label{tol1}
\end{eqnarray}
which leads to Tolman's relation between local temperature and lapse function
\begin{eqnarray}
L T = T_{\infty}, \label{tolman}
\end{eqnarray}
where $T_{\infty}$ is    the temperature  seen by an observer at infinity (where $L = 1$).

\subsection{The free-entropy representation: examples}
The above analysis demonstrates that  gravitating fluids are best described in the free-entropy representation. In this representation, the fundamental thermodynamic quantities depend only on the energy density $\rho$ and the activities $b_a$, and the latter are constant for entropy-maximizing solutions. As the temperature $T$ has a simple relation to the lapse function $L$, it is convenient  solve the equation $ T^{-1} = (\partial \omega/\partial \rho)_{b_a}$ for $T$, to express the energy density $\rho$, the pressure $P$ and the number densities $n_a$ as functions of $T$ and $b_a$.

  The standard textbook treatment of free fermion and boson gases   leads to expressions for $\rho, P$ and $n$ as a function of $T$ and $b$ \cite{huang},
\begin{eqnarray}
n(T, b) &=& \frac{g}{\pi^2} \int_0^{\infty} \frac{dp p^2}{e^{- b +  \epsilon_p/T} \pm 1} \\
P(T, b) &=&  \frac{g T}{\pi^2} \int_0^{\infty} dp p^2 \log\left[ 1 \pm e^{b - \epsilon_p/T}\right],
\end{eqnarray}
where $\epsilon_p$ stands for a particle's energy as a function of the momentum $p$, $+$ applies to fermions, $-$ to bosons and $g$ is the spin degeneracy. The density  $\rho$ is obtained by $\rho = T (\partial P/\partial T)_b  - P$. Note that in this example there is only one type of fermion, so the index $a$ is dropped.

For ultra-relativistic particles ($\epsilon_p = p$) with  $g = 2$,
\begin{eqnarray}
n(T, b) = -\frac{2}{\pi^2}f^{\pm}_1(b) T^3 \\
P(T, b) = -\frac{2}{\pi^2} f^{\pm}_2(b) T^4,
\end{eqnarray}
where $f_1^+ = -\mbox{Li}_3(-e^b)$, $f_1^- = \mbox{Li}_3(e^b)$, $f_2^+ = -\mbox{Li}_4(-e^b)$, and $f_2^- = \mbox{Li}_4(e^b)$;  $\mbox{Li}_k(x) := \sum_{n=1}\frac{x^n}{n^k}$ is the polylogarithm.

 It is then straightforward to derive the equation of state
 $\rho = 3 P$ and the free entropy functional
\begin{eqnarray}
\omega(\rho, b) = \frac{8}{6^{3/4}\sqrt{\pi}} \rho^{3/4} [f^{\pm}_2(b)]^{1/4}. \label{omurel}
\end{eqnarray}
For photons, the particle numbers are not preserved, hence, we set $b = 0$ in Eq. (\ref{omurel}). Note that the ultra-relativistic limit is identical to the limit $T \rightarrow \infty$.

Of interest is also the case of cold dilute gases, which correspond to constant $b$ and $T< < m$, where $m$ is the particle mass. In this regime, $\epsilon_p \simeq m+ \frac{p^2}{2m}$, and for $g = 2$,
\begin{eqnarray}
n(T, b) = \frac{1}{2} \left(\frac{2m}{\pi}\right)^{3/2} e^{b-\frac{m}{T}} T^{3/2},\label{ncold}\\
P(T, b) = \frac{1}{2} \left(\frac{2m}{\pi}\right)^{3/2}  e^{b-\frac{m}{T}} T^{5/2}, \label{pcold}
\end{eqnarray}
We recover the ideal gas EoS: $P = n T$, and
\begin{eqnarray}
\rho = (m + \frac{3}{2}T)n. \label{rcold}
\end{eqnarray}
The ideal gases have a thermodynamically consistent behavior at both limits $T\rightarrow 0$ and $T \rightarrow \infty$, according to the criterion that will be presented in Sec. 2.4.

\subsection{Thermodynamic inequalities}
Thermodynamic variables  are subject to constraints due to energy conditions and the requirement of  thermodynamic stability.

First, a standard thermodynamic assumption  is that the energy density $\rho$, the pressure $P$ and the temperature $T$ are positive. It follows that $\omega = (\rho +P)/T \geq 0$. Hence,
\begin{eqnarray}
(\partial P/\partial T)_{b_a} = \omega \geq 0.
\end{eqnarray}
In equilibrium,  the activities $b_a$ are constant. Hence,  pressure is an increasing function of temperature.

Thermodynamic stability implies that entropy function $s(\rho, n_a)$ is concave with respect to all  arguments. Its Legendre transform $\omega(\rho, b_a)$ is concave with respect to $\rho$ and convex with respect to $b_a$. It follows that $\left(\frac{\partial^2 \rho}{\partial \omega^2}\right)_{b_a} \geq 0 $. Since $\left(\frac{\partial^2 \rho}{\partial \omega^2}\right)_{b_a} = \left(\frac{\partial T}{\partial \rho}\right)_{b_a}$, we  conclude that
\begin{eqnarray}
(\partial \rho/\partial T)_{b_a}  \geq 0.
\end{eqnarray}
Hence, energy density is an increasing function of temperature.

We  also assume that the fluid satisfies the dominant energy condition, which implies that $P \leq \rho$. Since $\rho = \omega T - P$, we obtain $2P \leq \omega T = T \left( \frac{\partial P}{\partial  T}\right)_{b_a}$. This inequality has the trivial solution $P = 0, \rho = f(b_a) T, \omega = f(b_a)$, where $f$ is a function of the activities $b_a$. For non-zero pressure, it implies that
\begin{eqnarray}
\kappa := \left( \frac{\partial \log P}{\partial \log T}\right)_{b_a} \geq 2 \label{kappa1}
\end{eqnarray}
Hence, both the pressure $P$ and the energy density $\rho$ grow at least with $T^2$.
Note that the weak energy condition $\rho \geq 0$ implies the weaker inequality $\kappa \geq 1$.

 Eq. (\ref{kappa1}) implies that
 \begin{eqnarray}
 \lim_{T \rightarrow \infty} P(T, b_a) = \lim_{T \rightarrow \infty} \rho(T, b_a) = \infty.
 \end{eqnarray}
   In the limit $T \rightarrow 0$ with $b_a$ fixed\footnote{The usual limit $T \rightarrow 0$ that is employed in textbooks when treating phenomena like fermion degeneracy pressure or Bose-Einstein condensates is taken with $n_a$ constant (rather than $b_a$ constant).}, the chemical potentials $\mu_a = b_a T$ vanish. Hence, by   Eq. (\ref{euler}) $\rho + P \rightarrow 0 $. Since both $\rho$ and $P$ are non-negative, we conclude that
 \begin{eqnarray}
 \lim_{T \rightarrow 0} P(T, b_a) = \lim_{T \rightarrow 0} \rho(T, b_a) = 0.
 \end{eqnarray}
 Eq. (\ref{kappa1}) implies that both $\rho$ and $P$ drop at least as fast as $T^2$ as $T \rightarrow 0$.

\subsection{Thermodynamic consistency}
The Einstein equations for gravitating matter in equilibrium form a closed system of equations if a functional relation between pressure $P$ and energy density $\rho$ is specified. This relation is usually referred to as an equation of state. Indeed, thermodynamics predict a functional relation between $P$ and $\rho$ for constant $b_a$. However, not all functional relations of this form are thermodynamically consistent.

For fixed $b_a$, let $P = f(\rho)$ for some differentiable function $f: {\pmb R}_*^+ \rightarrow {\pmb R}_*^+$. By Eq.  (\ref{dpt}),
\begin{eqnarray}
\frac{d \rho}{dt} = \frac{\rho+f(\rho)}{f'(\rho)}, \label{tds}
\end{eqnarray}
 where $t = \ln T$. The universality of temperature implies that in thermodynamic systems, temperature can take any value in  ${\pmb R}^+$. Hence, the dynamical system (\ref{tds}) must admit a smooth solution for all $t \in (-\infty, \infty)$.  Then, the following criterion of thermodynamic consistency of an EoS follows.

 \bigskip

 \noindent {\em Thermodynamic Integrability.} An EoS $P = f(\rho)$ is thermodynamically consistent, if and only if the vector field  $X_f := \frac{x+f(x)}{f'(x)} \frac{\partial}{\partial x}$ on ${\pmb R}_*^+$ is complete.

\bigskip
An incomplete vector field $X_f$ may lead  to either infinite or zero density $\rho$ for finite temperature $T$.

For example, consider a function $f(x)$ with asymptotic behavior $f(x) = k x^a$ as $x \rightarrow \infty$, for $k, a >0$. By the dominant energy condition, $a \leq 1$. Take $a < 1$. For sufficiently large $\rho$, Eq. (\ref{tds}) implies that $\frac{d \rho}{dt} = \frac{\rho^{2-a}}{ka}$. Integrating from a point with density $\rho_0$ and temperature $T_0$, we find
\begin{eqnarray}
\frac{1}{\rho^{1-a}} = \frac{1}{\rho_0^{1-a}} - \frac{1-a}{ak} \log(T/T_0).
\end{eqnarray}
We note that   $\rho \rightarrow \infty$, for $T  = T_0 \exp\left( \frac{ak}{(1-a)\rho_0^{1-a}} \right)$. Only the exponent $a = 1$ is viable for the asymptotic behavior of $f$ at large $\rho$.

Similarly, consider  a function $f(x)$ such that $f(x) = k' x^{a'}$ as $x \rightarrow 0$, for $k', a' >0$. By the dominant energy condition, $a' \geq 1$. Take $a' > 1$. For sufficiently small $\rho$,  Eq. (\ref{tds}) implies that $\frac{d \rho}{dt} = (k'a'\rho^{a'-2})^{-1}$. Integrating from a point with density $\rho_0$ and temperature $T_0$, we find
\begin{eqnarray}
\rho^{a'-1} = \rho_0^{a'-1} +\frac{a'-1}{a'k'} \log(T/T_0).
\end{eqnarray}
We find that $\rho = 0$ for  $T  = T_0 \exp\left( -\frac{a'k'\rho_0^{a'-1}}{a'-1} \right)$. Only the exponent $a' = 1$ is viable for the asymptotic behavior of $f$ at small $\rho$.

\section{Characterization of solutions to the TOV equation}

\subsection{Preliminaries}
Consider a static spherically symmetric geometry
\begin{eqnarray}
ds^2  = - L^2(r) dt^2 + \frac{dr^2}{1-\frac{2m(r)}{r}} + r^2 (d\theta^2 +  \sin^2\theta d \phi^2) \hspace{0.2cm} \label{metric}
\end{eqnarray}
where $L(r)$ is the lapse function, $m(r)$  is the mass function, and $(t, r, \theta, \phi)$ is the adapted coordinate system.

Einstein's equations for the metric (\ref{metric}) lead to  the   Tolman-Oppenheimer-Volkoff (TOV) equation
\begin{eqnarray}
\frac{dP}{dr}  = - \frac{(\rho +P) (m+4\pi r^3P)}{r^2(1-\frac{2m}{r})}, \label{TOV}
\end{eqnarray}
where $\rho$ is the density and $P$ is the pressure  of matter. The TOV equation is supplemented by an equation for the mass function
\begin{eqnarray}
\frac{d m}{dr} = 4 \pi r^2 \rho, \label{dm}
\end{eqnarray}
while Eq. (\ref{cont}) becomes
\begin{eqnarray}
\frac{1}{L} \frac{dL}{dr} = -\frac{1}{\rho +P} \frac{dP}{dr} \label{dN}.
\end{eqnarray}
The set of equations (\ref{TOV}) and (\ref{dm}) is closed if an EoS that related pressure $P$ and energy density $\rho$ is specified. In what follows, we will assume that the EoS is thermodynamically consistent, in the sense that it satisfies the integrability condition of Sec. 2.4.

A thermodynamically consistent description of matter implies that the pressure $P$ and the energy density $\rho$ are functions of temperature $T$ and the activities $b_a$: $\rho(T, b_a)$ and $P(T, b_a)$. These functions are defined for all temperatures $T$. Since the activities $b_a$ are constant in equilibrium, the only independent variables in the set of equations (\ref{TOV}) and (\ref{dm}) is the temperature and the mass function. Using Eq. (\ref{dpt}), Eqs.  (\ref{TOV}, \ref{dm}) can be written equivalently as
\begin{eqnarray}
 \frac{d \log T}{dr} &=& - \frac{m + 4 \pi r^3 P(T)}{r^2 (1 - \frac{2m}{r})} \label{TOV2} \\
 \frac{dm}{dr} &=& 4 \pi r^2 \rho(T) \label{dm2}
\end{eqnarray}

Eqs. (\ref{TOV2}) and (\ref{dm2}) are to be integrated from a boundary point   $r = r_B > 0$ inwards, i.e., for $r < r_B$. Hence, they define an initial value problem with $M := m(r_B) <  2 r_B$ and $T_B := T(r_B) > 0$.   For any given $r_B$, the general solution to Eqs. (\ref{TOV2}) and (\ref{dm2}) is characterized by the $k+2$ parameters $(M, T_B, b_a)$ (recall, $k$ is the number of particle species). We assume that the functions $\rho(T, b_a)$ and $P(T, b_a)$ satisfy all thermodynamic inequalities of Sec. 2.3.

It is convenient to assume that for $r > r_B$, the metric corresponds to a Schwarzschild solution of mass $M$. The surface $r = r_B$ then places the role of a bounding box. The usual stellar boundary conditions correspond to $P = 0$. However, the condition $P << M/r_B^3$ is   physically sufficient for a stellar surface.

It is convenient to introduce the variables
\begin{eqnarray}
u :&=& \frac{2m}{r} \\
v :&=& 4 \pi r^2 \rho \\
w :&=& 4 \pi r^2 P \leq v \\
\xi :&=& - \log(r/r_B)\\
t :&=& \log(T/T_B).
\end{eqnarray}

Eqs. (\ref{TOV2}) and (\ref{dm2}) become
\begin{eqnarray}
\frac{dt}{d\xi} = \frac{\frac{1}{2}u+w}{1-u} \label{TOV3}\\
\frac{du}{d\xi} = u - 2v. \label{dm3}
\end{eqnarray}
The equations above are integrated from $\xi = 0$ to $\xi \rightarrow \infty$ ($r = 0$), with initial conditions $u(0) = u_B = \frac{2M}{r_B}$ and $t(0) = 0$.

In the remaining of the section, we will consider solutions to the above equations, with the stated boundary conditions. We will prove the following theorem.

\bigskip

\noindent {\em Theorem 1.} Integration of the TOV equations from the boundary inwards for a thermodynamically consistent EoS  proceeds all the way to the center. There are two types of solutions: regular ones ($m(0) = 0$) and singular ones with $\lim_{r \rightarrow 0} m(r) < 0$. For an EoS that satisfies $\lim_{T \rightarrow \infty} P/\rho = \lambda \leq 1$, singular solutions have finite $m(0)< 0 $ and temperature that vanishes with $r^{\frac{1}{2}}$ as $r \rightarrow 0$.

\subsection{Absence of horizons}

\noindent {\em Lemma 1.} The function $u(\xi)$ satisfies $u(\xi) < 1$ for all $\xi \geq 0$.

\medskip

\noindent{\em Proof.} Since $u(0) < 1$, assume for contradiction that $u(\xi)$ first becomes unity   at some point $\xi = \xi_* >0$. For $\xi < \xi_*$, $u$ is at least a
 $C^1$ function of $\xi$, and $\epsilon:= 1 - u > 0$. For $x:= \xi_* - \xi$ sufficiently small, Eqs.  (\ref{TOV2}) and (\ref{dm2}) become
 \begin{eqnarray}
 \frac{dt}{dx} = -\frac{\frac{1}{2}+w}{\epsilon}  \label{TOV4}\\
\frac{d\epsilon}{dx} = 1 - 2v. \label{dm4}
 \end{eqnarray}
To reach $\epsilon = 0$, $\frac{d\epsilon}{dx} \geq 0$ as $ x\rightarrow 0_+$, hence, $1-2v \geq 0$. It follows that $v \leq \frac{1}{2}$ and hence, $w \leq \frac{1}{2}$.

If $v_*= v(\xi_*) < \frac{1}{2}$, then Eq. (\ref{dm4}) implies that $\epsilon$ is well approximated by $(1 - 2v_*)x$ in the vicinity of $x = 0$. Hence,  there is a referennce point $x_r$ and a constant $C > 0$ such that $\epsilon < Cx$, for all $x \in[0, x_r)$.
Eq. (\ref{TOV}) becomes
$- \frac{dt}{dx} < \frac{1}{\epsilon} < \frac{1}{Cx}$. Integrating both terms from  $x_r$ to $x$, we obtain
\begin{eqnarray}
- \log \frac{T(x)}{T(x_r)} < C^{-1} \log(x/x_r) .
\end{eqnarray}
It follows that $T(x)$ diverges as $x \rightarrow 0^+$ with $x^{-C}$. This implies that $\rho$ also diverges at $x = 0$, hence, so does $v$, contradicting the condition $v \leq \frac{1}{2}$.

For $v_* = \frac{1}{2}$, we write $ 1-2v = f(x)$ for some function $f$ that vanishes for $x \rightarrow 0$. Hence, $\frac{d\epsilon}{dx} = f(x)$. Hence $\epsilon = F(x) = \int_0^xdx'f(x')$ and $F(x)$ is a function that vanishes faster than $x$ as $x \rightarrow 0^+$. Integrating from a reference point $x_r$ to $x$, we find  $- \log \frac{T(x)}{T(x_r)} < \int_{x_r}^x \frac{dx'}{F(x')}$, and $T$ again diverges as $x \rightarrow 0^+$.

Thus, we demonstrated that the assumption $u(\xi_*) = 1$ leads  to contradiction. ${\small \blacksquare}$

\bigskip

Hence, no horizon is encountered when integrating the TOV from the boundary inwards. However, in some cases the function $u$ may take values very close to unity.
For example, in  self-gravitating radiation ($P = \frac{1}{3}\rho \sim T^4$), $u$ achieves a maximum value $u_{max} = 1 - \epsilon$, where $\epsilon$ is approximately proportional to $u_B \sqrt{v(0)}$. Hence, $\epsilon$ can become arbitrarily small by choosing sufficiently small boundary pressure as initial condition \cite{AnSav16}. These "approximate horizon" solutions can be used in order to model a black hole in a box, at equilibrium with its Hawking radiation.

Lemma 1 relies crucially on the positivity of pressure. For sufficiently large negative pressure, the temperature would not blow up on the horizon and the TOV would be consistent with the presence of horizon. Some examples are described in Ref. \cite{AnSav16}.

A trivial corollary of Lemma 1 is that for any finite interval $[0, \xi]$, there is a maximal value $u_m$ of $u$, such that $u_m < 1$.

Lemma 1 implies that $m(r) < \frac{1}{2} r$ for all $r < r_B$. Hence, $m(r)$ cannot take positive values as $r \rightarrow 0$. Hence, either $\lim_{r \rightarrow 0}m(r) = 0$ or $\lim_{r \rightarrow 0}m(r) < 0$. The former case corresponds to regular solutions; the latter case corresponds to singular solutions.




\subsection{Singular solutions: vanishing of the mass function}
The regular solutions to the TOV equation have been exhaustively studied in the literature, and we will not consider them in this paper. We   remind the reader that for any given EoS, there is a mass $M_{OV}$, the Oppenheimer-Volkoff limit,
such that there are no regular solutions with $M > M_{OV}$.  There exists also  EoS-independent bounds to $u_B$ for regular solutions \cite{Buchdahl, Buchother, Dadh}, like the Buchdahl limit, $u_B \leq \frac{8}{9}$.

Singular solutions
are defined by the condition $\lim_{r \rightarrow 0}m(r) < 0$, and they exist for all values of $M > 0$ and of $u_B \in (0, 1)$.
Since $m(r_B) = M >0$, continuity implies that
singular solutions are characterized by a radius  $0 < r_1 < r_B$, such that $m(r_1) = 0$. When integrating from the boundary inwards, we will encounter negative values of the mass function,  unless integration is interrupted first by a singularity.
In Ref. \cite{ST}, it was proven that no such singularities appear for positive $m$, hence,
integration from the boundary inwards always encounters a point $r_1 \geq 0$, where $m(r_1) = 0$. Ref. \cite{ST} then uses this result, in order to prove our Lemma 1.
The proof of Ref. \cite{ST} is more general than ours, in that it does not assume  thermodynamic consistency for the EoS; the restrictions to the EoS are much milder.

The assumption of thermodynamic consistency allows for a simpler proof, following directly from the Picard-Lindel\"of theorem \cite{DE} for the local existence and uniqueness of  solutions to ordinary differential equations.

\bigskip

\noindent {\em Lemma 2.} Integration from the boundary inwards  encounters a point $r_1 \geq 0$, where $m(r_1) = 0$.

\medskip

\noindent{\em Proof.}
Consider an interval $[r_0, r_B]$ with $r_0 > 0$. Let $u_m < 1$ be the maximum value of $u$ in this interval.  When writing the system of Eqs. (\ref{TOV2}) and (\ref{dm2}) as
\begin{eqnarray}
\frac{d t}{dr} = f_1(r, m, t) \hspace{1cm} \frac{dm}{dr} = f_2(r, m, t),
\end{eqnarray}
we note that the domain of the functions $f_1$ and $f_2$  excludes the singular points $r = 0$ (by definition) and $2m/r = 1$ (by Lemma 1). If $P$ and $\rho$ are differentiable functions of the temperature $T$, then $f_1$ and $f_2$ are differentiable functions of $r, m$ and $t$. They satisfy the conditions of the Picard-Lindel\"of theorem starting from any $r \in [r_0, r_B]$. Since $r_0$ can be brought arbitrarily close to $0$, the solution  connects either to a regular or a singular solution. In the former case $r_1 = 0$, in the latter case $r_1 > 0$. ${\small \blacksquare}$

 \medskip

\noindent {\em Corollary 1.} If $r_1 > 0$, then $T(r_1)$ is finite and non-zero.

\medskip

\noindent{\em Proof.} Since $dT/dr < 0$ for $m \geq 0$, $T(r_1) > T_R > 0$. Since $r_1$ is a regular point of the system, $T$ does not diverge there. ${\small \blacksquare}$

\medskip

Evidently, $P(r_1)$ and $\rho(r_1)$ are also finite and non-zero.

We also note that there is a point $r_0  \in [r_1, r_B]$, where $u$ takes its maximal value. If $du/dr(r_B) < 0 $, then $r_1$ corresponds to a local maximum of $u$. If $du/dr(r_B) > 0 $, then $u$ is decreasing in $[r_1, r_B]$, and $r_0$ coincides with $r_B$.

\subsection{Non-monotonicity of temperature}
Consider a solution with $m(r_1)$ = 0 for $r_1 > 0$, or equivalently $u(\xi_1) = 0$, where $\xi = - \log(r_1/r_B)$. Since $du/d\xi < 0$ for $u < 0$, $u(\xi)< 0$ for all $\xi > \xi_1$.

By Eq. (\ref{dm3}), $\frac{dt}{d\xi }(\xi_1) = w(\xi_1)> 0$, i.e., the temperature increases towards the center. However, this property does not extend to all $\xi > \xi_1$.

\bigskip

\noindent {\em Lemma  3.} For any solution with bounded mass function, there exists $\xi_2 > \xi_1$, such that $\frac{dt}{d\xi}(\xi_2) = 0$.

\medskip

\noindent{\em Proof.} By contradiction, if $\frac{dt}{d \xi} > 0$ for all $\xi > \xi_1$, then $w > -\frac{1}{2}u$ for all $\xi > \xi_1$. This implies that $v > -\frac{1}{2}u$, hence, $u-2v < 2u$. Then,
Eq. (\ref{dm3}) becomes $\frac{du}{d\xi} < 2 u$, which implies that $u < - k e^{2\xi}$, for some $k > 0$. This implies that the mass function $m(r)$ grows at least with $r^{-1}$ as $r \rightarrow 0$, hence, it is not bounded.
${\small \blacksquare}$

\bigskip

In solutions with unbounded mass function, $w$ and $v$ diverge faster than $r^{-2}$ at the center, hence, energy density $\rho$ and pressure $P$ diverging faster than $r^{-4}$.
Such solutions are unphysical but Lemma 3 does not forbid them. The key point here is that these solutions strongly depend on the behavior of the EoS as $T \rightarrow \infty$. If we constrain the asymptotic behavior of the allowed EoS, then these solutions disappear.

We will assume that the EoS satisfies
 \begin{eqnarray}
 \lim_{T \rightarrow \infty} \frac{P}{\rho} = \lambda \leq 1. \label{limeos}
  \end{eqnarray}
  An asymptotic behavior of this form is well justified by our analysis of the integrability condition in Sec. 2.4. The natural choice for $\lambda$ is $\frac{1}{3}$, which expresses the hypothesis that  for sufficiently high energy all particles behave like massless particles even in presence of interactions\footnote{ This discussion relates to an old problem, namely, what  conditions an EoS must satisfy in order to be compatible with relativity.  The causality condition $|(\partial P/\partial \rho)_s| \leq 1$ is well accepted, since it  guarantees that the speed of sound on the material never exceeds the speed of light.  Other conditions have been suggested but they are not universally accepted. For example, Landau and Lifschitz proposed \cite{LL} that $P \leq \frac{1}{3} \rho$.
 Counterexamples exist \cite{Zeld, GH}; however,   theories characterized by asymptotic freedom are expected to saturate the Landau-Lifschitz condition for $\rho \rightarrow \infty$.}.  The limiting value $\lambda = 1$ corresponds to the stiffest equation of state proposed by Zel'dovitch \cite{Zeld2}

 By Eq. (\ref{dpt}), we find  that asymptotically
 \begin{eqnarray}
 P   = \lambda \rho \sim T^{1+\frac{1}{\lambda}} \label{asymeos}
 \end{eqnarray}

 \noindent {\em Lemma  4.} Eq. (\ref{limeos}) disallows solutions with unbounded mass function.

\medskip

\noindent{\em Proof.}
Consider Eqs. (\ref{dm2}) and (\ref{TOV2}) as $r \rightarrow 0$. For solutions with unbounded negative $m$,
 \begin{eqnarray}
 \frac{d\log T}{d r} = \frac{m+4\pi r^3 P}{2mr} = \frac{1}{2r} + \frac{\lambda}{2mr}   \frac{dm}{dr} = \frac{1}{2r} + \frac{\lambda}{2r} \frac{d \log |m|}{dr}.
 \end{eqnarray}
 Integrating we find that $T |m|^{-\lambda/2} \sim \sqrt{r}$. By Eq. (\ref{dm2}), the energy density $\rho$ is proportional to $r^{-2} dm/dr$. Eq. (\ref{asymeos}) then implies that
 \begin{eqnarray}
 d|m|/dr = -K r^2 \left(r^{1/2}|m|^{\lambda/2}\right)^{1+ \frac{1}{\lambda}} = - K r^{\frac{5}{2} + \frac{1}{2\lambda}}|m|^{\frac{1}{2} + \frac{\lambda}{2}}, \label{dmrr}
 \end{eqnarray}
for some $K> 0$.

For $\lambda < 1$, the general solution to Eq. (\ref{dmrr})  is   $|m|^{(1-\lambda)/2} = c_1 - c_2 r^{\frac{7}{2} + \frac{1}{2\lambda}}$, for $c_1, c_2 \geq 0$. This solution is bounded as $r \rightarrow 0$, in contradiction to the hypothesis.

In the limiting case $\lambda = 1$, the solution to Eq. (\ref{dmrr}) is $|m| = C e^{-\frac{Kr^5}{5}}$, where $C > 0$. This solution is also bounded as $r \rightarrow 0$, also contradicting the hypothesis.
 ${\small \blacksquare}$

\bigskip

The condition  (\ref{limeos}) does not constrain physics,
because  it refers solely to asymptotic properties of the EoS. Any known physical EoS is valid up to a maximum temperature $T_m$. To remove solutions with unbounded mass function, it suffices that the extrapolation of the EoS to temperatures $T > T_m$ satisfies   Eq. (\ref{limeos}).

\subsection{Integration to the center}
We assume an EoS that satisfies Eq. (\ref{limeos}). Hence, there exists a point $\xi_2$ where $dt/d\xi$ vanishes, or equivalently $u_2:=u(\xi_2) = - 2 w(\xi_2)$.

\medskip

 \noindent {\em Lemma 5.} $dt/d\xi <0$ for all $\xi > \xi_2$.

\medskip

\noindent{\em Proof.} We calculate
\begin{eqnarray}
\frac{d^2t}{d \xi^2} = \frac{(\frac{1}{2}+w)}{(1-u)^2} \frac{du}{d\xi} - \frac{2w}{1-u} + \kappa \frac{w}{1-u} \frac{dt}{d\xi}, \label{2dtdx}
\end{eqnarray}
where $\kappa$ is given by Eq. (\ref{kappa1}).
For $u < 0$, the first two terms in the right-hand side of Eq. (\ref{2dtdx}) are negative. This implies that $\frac{d^2t}{d \xi^2} (\xi_2) < 0$, hence, $\xi_2$ is a local maximum of $t(\xi)$. By continuity, there exists a point $\bar{\xi}_2 > \xi_2$, such that $\frac{dt}{d\xi}(\bar{\xi}_2 ) < 0$ for all $\xi \in(\xi_2, \bar{\xi}_2]$.

Eq. (\ref{2dtdx}) implies that
\begin{eqnarray}
\frac{dy}{d\xi} > g(\xi) y, \label{dydx}
\end{eqnarray}
for $y = -\frac{dt}{d\xi}$ and   some non-negative function $g(\xi)$.

We integrate Eq. (\ref{dydx}) from $\xi_2'$ to any $\xi > \bar{\xi}_2 $, to obtain $y(\xi)/y(\bar{\xi}_2 ) = exp[\int_{\bar{\xi}_2 }^{\xi} d\xi' g(\xi')] > 0$. Hence $\frac{dt}{d\xi}(\xi) < 0$. ${\small \blacksquare}$

 \medskip

 Integration then proceeds smoothly to all $\xi > \xi_2$.

 \bigskip

 \noindent {\em Lemma 6.} As $\xi \rightarrow \infty$, $t(\xi) \sim -\frac{1}{2}\xi$ and $u(\xi) \sim e^{\xi}$.

\medskip

\noindent{\em Proof.} Integrating the inequality $du/d\xi < u$ from $\xi_2$ to any $\xi > \xi_2$, we obtain $u(\xi) < u_2 e^{\xi}$. We note that
\begin{eqnarray}
0 < \frac{dt}{d\xi} +\frac{1}{2} = \frac{\frac{1}{2}+w}{1-u} < \frac{\frac{1}{2}+w(\xi_2)}{1 - u} <  \frac{\frac{1}{2}+w(\xi_2)}{- u} = \frac{1}{2}(1+|u_2|^{-1})e^{-\xi}.
\end{eqnarray}
It follows that $dt/d\xi \rightarrow -\frac{1}{2}$ as $\xi \rightarrow \infty$, and $t \sim   -\frac{1}{2}\xi$.  This implies that $\rho$ vanishes as $\xi \rightarrow \infty$, hence, $v$ drops to zero faster than $e^{-2\xi}$. By Eq. (\ref{dm3}),  $d\ln u/d\xi \rightarrow 1$, hence,  $u(\xi) \sim e^{\xi}$.
 ${\small \blacksquare}$

\bigskip

Hence,   $m(r)$ tends to a negative constant as $r \rightarrow 0$ and $T(r)$ vanishes with $\sqrt{r}$. By Tolman's law, the lapse function diverges with $r^{-1/2}$. Theorem 1 has been proven.

\begin{figure}[]
\includegraphics[height=7cm]{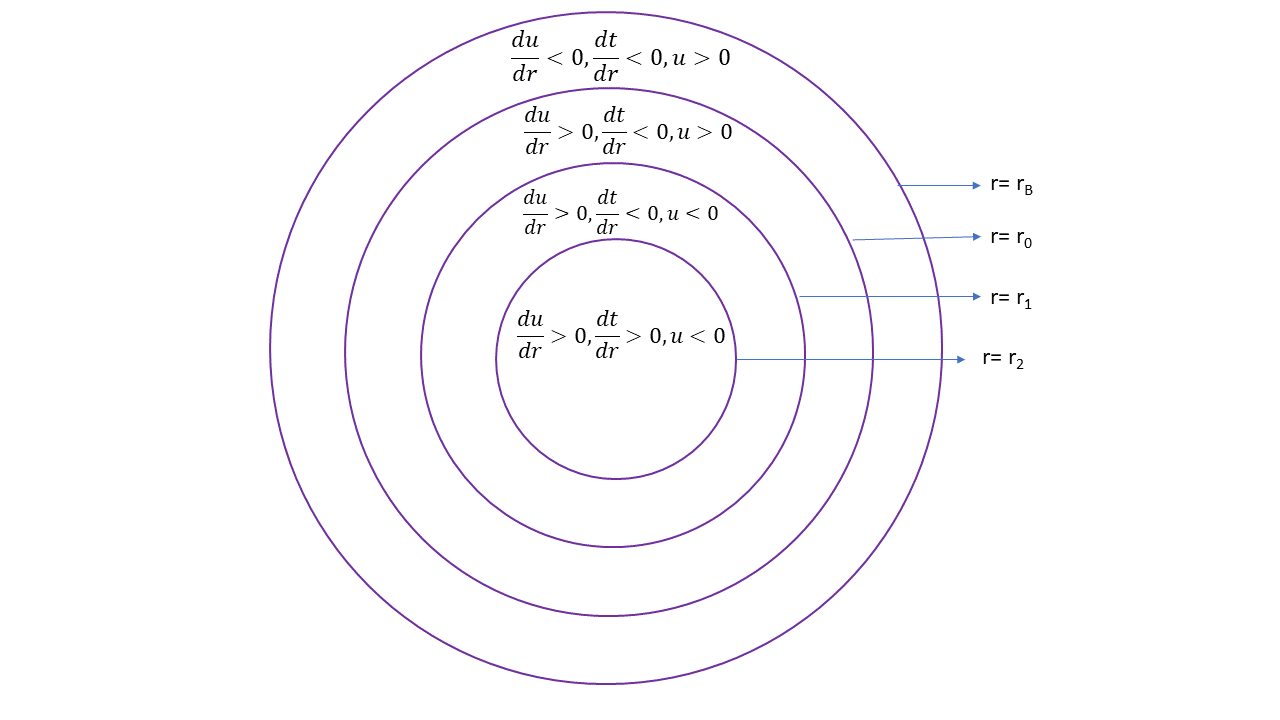} \caption{ \small     The structure of a singular solution to the TOV equation. Integration starts at $r = r_B$ and proceeds inwards. At the point $r = r_0$, the function $u$ is maximized (in solutions with $\frac{du}{dr}(r_B) > 0$, $r_0$ coincides with $r_B)$. At $r = r_1$, the mass function $m(r)$ vanishes, and so does $u = 2m/r$. At $r = r_2$ temperature is maximized. Temperature vanishes at the center.
}\label{structure}
\end{figure}

\begin{figure}[]
\includegraphics[height=6cm]{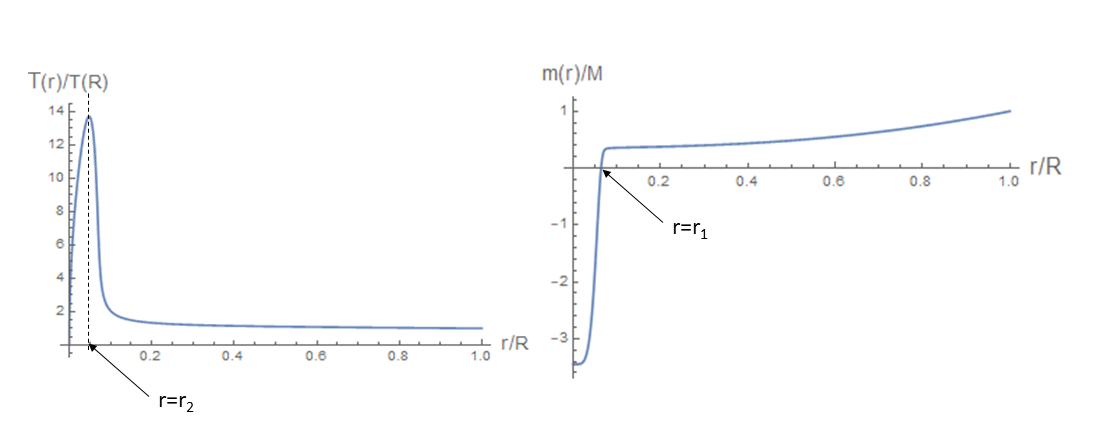} \caption{ \small     The temperature and a mass function for a representative singular solution to the TOV equation for $P = \frac{1}{3}\rho$. This solution is obtained for $u_R = 0.2$ and $v_R$. The points $r_1$ and $r_2$ are indicated. For this solution $r_1/R \simeq 0.064$ and $r_2/R \simeq 0.048$.
}\label{example}
\end{figure}

To summarize, for any thermodynamically consistent EoS subject to the asymptotic condition (\ref{limeos}), the TOV equations can be integrated from the boundary inwards to $r = 0$. There are two types of solutions: (i) regular, with $m(0) = 0$, and singular with $m(0) = - M_0 < 0$. All singular solutions are characterized by a point $r_1 < r_B$ at which the mass function vanishes, and by a point $r_2 < r_1$ that is a local maximum of temperature. Temperature (and hence, density and pressure) decreases for $r < r_2$ and vanishes with $\sqrt{r}$ as $r \rightarrow 0$. A schematic representation of the structure of a singular solution is given in Fig. \ref{structure}, while a plot of a representative solution is shown in Fig. \ref{example}. 

In the next section, we will discuss properties of the singular solution in more detail, emphasizing in particular the nature of the singularity at $r = 0$.

\section{Properties of singular solutions}

\subsection{Relation to the negative-mass Schwarzschild singularity}

As shown in the previous section, a singular solution to the TOV equations is characterized by $m(0) = - M_0$ for   positive   $M_0$ and by a lapse function that diverges as
\begin{eqnarray}
L (r) = \frac{\eta}{\sqrt{r}},
\end{eqnarray}
for some constant $\eta > 0$. The two parameters $M_0$ and $\eta$ fully characterize the structure of the solutions near $r = 0$.

Around $r = 0$, the metric (\ref{metric}) becomes
  \begin{eqnarray}
ds^2  = - \frac{\eta^2}{r} dt^2 + \frac{r dr^2}{2M_0} + r^2 (d\theta^2 + \sin^2\theta d \phi^2). \label{metric3}
\end{eqnarray}

The metric Eq. (\ref{metric3}) has the same asymptotic behavior with a Schwarzschild solution with negative mass $-M_0$, modulo a time rescaling. Indeed, Eq. (\ref{metric3}) can be expressed as
\begin{eqnarray}
ds^2  = - \frac{2M_0}{r} d\tilde{t}^2 + \frac{r dr^2}{2M_0} + r^2 (d\theta^2 + \sin^2\theta d \phi^2), \label{metric4}
\end{eqnarray}
where $\tilde{t} = \frac{\eta}{\sqrt{2M_0}}t $.

Next we evaluate the sub-leading terms to $g_{00}$ and $g_{rr}$ as $r \rightarrow 0$. Since $\rho$ vanishes at least with $T^2$ as $T \rightarrow 0$ with $b_a$ fixed, $\rho$ grows at most with $r$ near $r = 0$. By Eq. (\ref{dm}), $dm/dr$ vanishes at least with $r^3$. Hence, $m(r) = - M_0 + k_1 r^{\delta}$, where $k_1 > 0$ and $\delta \geq 4$.

It follows that the metric component
\begin{eqnarray}
g_{rr} = \frac{r}{2M_0} \left[ 1 - \frac{r}{2M_0} + \frac{r^2}{4M_0^2} - \frac{r^3}{M_0^3} + O[(r/M_0)^4]   \right]
\end{eqnarray}
is equal with the metric component  $(1+2M_0/r)^{-1}$ of a negative-mass Schwarzschild geometry, up to terms of order $(r/2M_0)^4$. The first matter-dependent term appears at order $(r/2M_0)^4$ or higher.

Eq. (\ref{TOV2}) implies that near $r = 0$
\begin{eqnarray}
r \frac{d \log T}{dr} = -\frac{-M_0+kr^{\delta}+4\pi r^3 P}{2M_0 +r -kr^{\delta}}. \label{TOV5}
\end{eqnarray}
Since $P(r)$ grows at most with $r$ near zero, Eq. (\ref{TOV5}) becomes
\begin{eqnarray}
r \frac{d \log T}{dr} = \frac{1}{2} (1 + \frac{r}{2M_0})^{-1},
\end{eqnarray}
up to terms of order at least $(r/M_0)^4$. Hence,
\begin{eqnarray}
g_{tt} = \frac{\eta^2}{2M_0} \left(1+ \frac{2M_0}{r} \right) + O[(r/M_0)^4].
\end{eqnarray}

This means that for $r << 2M_0$, the metric corresponds to a vacuum solution of Einstein' equation. There is little matter in the vicinity of the singularity.

The curvature around $r = 0$ can be calculated using the negative-mass Schwarzschild spacetime. This means that the Ricci tensor vanishes, while the Kretschmann scalar $K = R_{\mu\nu\rho \sigma} R^{\mu\nu\rho \sigma} $ is given by
\begin{eqnarray}
K:= \frac{48M_0^2}{r^6}.
\end{eqnarray}
Hence, $r = 0$ is a curvature singularity, and it is naked as it is not covered by any horizon.

We also note that the area to volume ratio of a small sphere of area $A$ around $r = 0$, decreases with $A^{-3/4}$, while the same quantity around a regular point decreases with $A^{-1/2}$.

\subsection{Geodesics near the singularity}

Next, we analyze the properties of the singularity at $r = 0$ in relation to  the causality and predictability properties of the spacetime. To this end, we study the geodesic equation
  near $r = 0$,
\begin{eqnarray}
\dot{r}^2 = \frac{2M_0\epsilon^2}{\eta^2} - \frac{2M_0\sigma}{r} - \frac{2M_0 \ell^2}{r^3}, \label{geodes}
\end{eqnarray}
where $\epsilon$ and $\ell$ are constants. For causal geodesics, $\epsilon$ correspond to energy per unit mass and it satisfies
\begin{eqnarray}
\dot{t} \eta^2 = \epsilon r; \label{dte}
\end{eqnarray}
$\ell$ is the angular momentum per unit mass, and it satisfies $r^2 \dot{\phi} = \ell$.  The parameter $\sigma$ takes the value $1$ for timelike, $-1$ for spacelike  and $0$ for null geodesics. The dot denotes derivative with respect to an affine parameter $\lambda$ that increases towards the future direction.

The singularity strongly repulses all test particles. No timelike geodesics  arrive at the singularity.  Incoming massive particles reach at most up to a minimal radius, $r_{min} = (\eta/\epsilon)^2$ (achieved for $\ell = 0$),  and then they bounce back. Hence, all timelike geodesics that start from past timelike infinity $\iota^-$ reach the future timelike infinity $\iota^+$.  The spacetime is timelike geodesically complete.



Null geodesics with $\ell \neq 0$ also reach a minimal radius $r_{min} = (\eta \ell/\epsilon)^{2/3}$. The only causal geodesics that reach the singularity are radial ($\ell = 0$) null geodesics, and these form a set of measure zero in the space of all null geodesics.
These satisfy $ \dot{r} = \pm \sqrt{2M_0}\epsilon/\eta $; hence, $r =  \pm (\sqrt{2M_0}\epsilon/\eta) \lambda$, for a path parameter $\lambda$ that vanishes at $r = 0$. By Eq. (\ref{dte}), $t = t_0 +  \frac{1}{2} (\sqrt{2M_0}\epsilon^2 \eta^3) \lambda^2$, for some constant $t_0$. The $+$ solution corresponds to outgoing geodesics and the $-$ solution to incoming geodesics.

\begin{figure}[] \label{geod1}
\includegraphics[height=7cm]{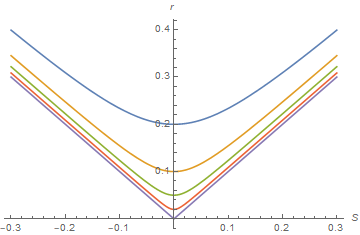} \caption{ \small  The radius coordinate $r$ as a function of the rescaled affine parameter $S = \sqrt{2M_0} \epsilon \lambda/\eta$ for the null geodesics of the metric (\ref{metric3}). The affine parameters are chosen so that $S= 0 $  for minimum $r$. Plotted geodesics differ on the value of the angular momentum $\ell$. From top to bottom, the value of the parameter $(\eta \ell/\epsilon)^{2/3}$ is $0.2, 0.1, 0.05, 0.02$ and $0$. The zero angular momentum geodesic is non-differentiable at $r = 0$.    }
\end{figure}

In a recent analysis of the singularity in negative-mass Schwarzschild spacetime \cite{SeQu}, the divergence of curvature at $r = 0$ was given as a justification for the incompleteness of these geodesics. This statement presupposes that the point $r = 0$ has been excised from the spacetime manifold. If, however, $r = 0$ is treated as a spacetime boundary or ideal point in the sense of Geroch, Kronheimer and Penrose \cite{GKP},  we can interpret $r = 0$ as the point where    incoming future directed geodesic becomes outcoming future directed geodesics\footnote{Contrast this situation with radial null geodesics in the positive-mass Schwarzschild spacetime. The only  future-directed geodesics around $r = 0$, are of the form $r = - \epsilon \lambda$, because $\frac{\partial}{\partial r}$ is a past-directed timelike vector field near $r = 0$. Incoming geodesics terminate at $r = 0$.}. The radial geodesics are continuous but non-differentiable at $r = 0$, and they can be defined as the limit of differentiable geodesics with nonzero angular momentum $\ell$, at the limit $\ell \rightarrow 0 $---see, Fig. \ref{geod1}.
 Hence, all null geodesics that start from the past null infinity $\mathcal{I}^-$ end at the future null infinity $\mathcal{I}^+$.

Note that null geodesic propagation through $r = 0$ takes place at finite time. Consider a static observer   at $r = r_0$ and $\phi =0$ on the equatorial plane who sends a light ray towards the center. Consider also a mirror at $r = r_0$ and $\phi = \pi$ that reflects the outgoing ray. The initial observer will detect the reflected  light ray after   {\em finite} proper time $\delta \tau = \frac{2 \eta}{\sqrt{2M_0}} r_0$  that vanishes as $r_0 \rightarrow 0$.

To summarize,   singular solutions to the TOV equations are causally complete. The only trouble at the level of causal geodesics is the non-differentiability of radial null geodesics at the singularity.
Hence, the singularity at $r = 0$ is much more benign than black hole singularities, despite its nakedness.

\subsection{Bounded acceleration paths}
Causal geodesic completeness is a minimal condition for a spacetime to be considered singularity-free \cite{HawkingEllis}. Of course, observers  are not necessarily free-falling, they may be accelerated. However, any spaceship moving towards the singularity can only have finite acceleration and it can carry only a finite amount of fuel. This implies that  the physically relevant criterion is the completeness of bounded-acceleration trajectories  \cite{Ger68}. We will study the singularity at $r = 0$ in relation to this criterion.

The acceleration one-form for static observers near the singularity is
\begin{eqnarray}
a = \frac{d \log L}{dr} dr = -\frac{1}{2r} dr. \label{accel}
\end{eqnarray}
The minus sign in Eq. (\ref{accel}) implies a repulsive force.  The proper acceleration $\sqrt{a^{\mu}a_{\mu}}$ diverges like $r^{-1/2}$  as $r \rightarrow 0$.  In static configurations,  infinite pressure is required in order to push a material element towards the singularity.

Next, we consider an infalling observer on a timelike curve with four-velocity
 $u^{\mu} = (\dot{t}, \dot{r}, \dot{\theta}, \dot{\phi})$. Since $u^{\mu}u_{\mu} = -1$,
\begin{eqnarray}
\dot{t} = \frac{\sqrt{r}}{\eta} \sqrt{1 + \frac{r\dot{r}^2}{2M_0} + r^2 \dot{\theta}^2 + r^2 \sin^2\theta \dot{\phi}^2}. \label{dott0}
\end{eqnarray}
If $\dot{r}, \dot{\theta}$ and $\dot{\phi}$ are bounded, then
\begin{eqnarray}
\dot{t} \simeq \frac{\sqrt{r}}{\eta} \label{dottrr}
 \end{eqnarray}
 as $r \rightarrow 0$. Hence, $u^{\mu}$ approximates the four-velocity of a static observer. By Eq. (\ref{accel}), the observer requires infinite acceleration to reach the singularity.

We examine the possibility that $\dot{r}$  diverges as $r \rightarrow 0$. Since the tangential acceleration does not affect whether the observer reaches the singularity or not, the divergence of $\dot{\theta}$ or $\dot{\phi}$ is irrelevant.

  Let $\dot{r}$ diverges with  $r^{-1/2}$ or more  slowly. Then,  Eq. (\ref{dottrr}) still applies and the earlier conclusion remains unchanged. If $\dot{r}$ diverges faster that $r^{-1/2}$, then by Eq. (\ref{dott0}),
\begin{eqnarray}
\dot{t} = \frac{r |\dot{r}|}{\eta \sqrt{2M_0}}. \label{accelr}
\end{eqnarray}
However, this expression is not compatible with bounded acceleration. Taking $\dot{r} \sim r^{-s}$ for $s > \frac{1}{2}$, we find that $a^2:= a_{\mu}a^{\mu} \sim r^{-(4s+1)}$, hence, divergent.

  We conclude that no observer with finite proper acceleration reaches the singularity. There are no incomplete timelike paths with bounded acceleration. The spacetime is bounded-acceleration complete.

This conclusion allows us to characterize the singularity at $r = 0$ as an ideal point \cite{GKP} in the following sense.  Let $I^-_b(p)$ be the set of all points $q$ in the past of $q$ along a timelike curve of bounded acceleration. Then, $r = 0$ can be identified with part of the boundary of $I^-_b(\iota^+)$, where $\iota^+$ is the future timelike infinity. This characterization is compatible with the characterization of $r = 0 $ as a conformal boundary, to be presented next.

\subsection{ The surface $r = 0$ as a conformal boundary}

Consider a singular conformal transformation $g \rightarrow \tilde{g}  := f^2 g$ of a singular solution to the TOV equations. We choose  $f = L^{-1}$. Around $r = 0$, the metric $\tilde{g}$  is
\begin{eqnarray}
d \tilde{s}^2 = - d\tilde{t}^2 + dx^2 +  (2x)^{3/2} (d\theta^2 + \sin^2\theta d \phi^2) \label{ultrastatic}
\end{eqnarray}
where $\tilde{t} = \eta^2/(8M_0^3) t$ and $x = \frac{r^2}{8M_0^2}$. The spacetime $\tilde{M}$ described by the metric  (\ref{ultrastatic}) has a   boundary $\partial M = {\pmb R} \times S^2$ at $x = 0$, and the pull-back of the four metric on $\partial M$ is a Lorentzian metric. Hence, $(\tilde{M}, \tilde{g})$ is a spacetime with timelike boundary \cite{HFS}.

The metric $\tilde{g}$ is ultrastatic. In ultrastatic spacetimes,  causal geodesics correspond to geodesics of the pull-backed three metric
\begin{eqnarray}
d \sigma^2 =  dx^2 +  (2x)^{3/2} (d\theta^2 + \sin^2\theta d \phi^2). \label{3dg}
\end{eqnarray}
on the spatial manifold $\tilde{\Sigma} = {\pmb R}^+ \times S^2$ that is  described by the local coordinates $(x, \theta, \phi)$.


$\tilde{\Sigma}$ is a manifold with boundary.  The boundary $\partial \tilde{\Sigma} = S^2$ is defined by $x = 0$.  A submanifold of constant $x$ is a two-sphere of area $\frac{4\pi}{5}(2x)^{5/2}$.  The solid angle of a sphere of proper radius $x$ around $x = 0$ goes to zero with $\frac{4\pi}{5} \sqrt{2x}$.

The three-metric (\ref{3dg}) has negative curvature near $x = 0$, and the Ricci scalar diverges as $-\frac{3}{8x^2}$ as $x \rightarrow 0$. Despite the divergence of the curvature, curves can be  continued across the singularity. In this sense, the boundary $x = 0$ behaves like a conical singularity.

To see this, we first note that the singularity at $x = 0$ is reached only by radial geodesics. These geodesics are complete, but they are not differentiable at the singularity. Furthermore, the proper distance of any point $(x, \theta, \phi)$ from the singularity is finite and equal to $x$.
We do not provide a proof of these statements, as they follow from an analysis that is almost identical to that of Sec. 4.2.

The key point is that the causal structure of an ultrastatic spacetime is Newtonian, with the time $\tilde{t}$ as a Newtonian time parameter. This is because the vector field $\frac{\partial}{\partial \tilde{t}}$ is covariantly constant. This implies that all continuous timelike paths can be parameterized by $\tilde{t}$, and that each (inextendible) path intersects a surface  of constant $\tilde{t}$ only once.  Hence, the spacetime $(\tilde{M}, \tilde{g})$ is a globally hyperbolic spacetime with boundary \cite{HFS}.

\begin{figure}[tbp]
\hspace{5cm}
\includegraphics[height=6.5cm]{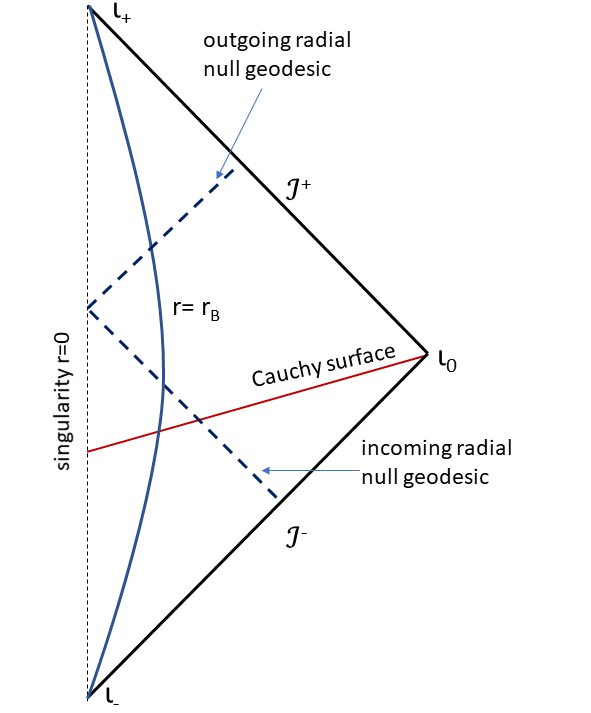} \caption{ \small   Penrose diagram for a singular solution to the TOV equations. } \label{Penrd}
\end{figure}

A Penrose diagram for the spacetime $(M, g)$ is given in Fig. \ref{Penrd}.

\section{Limiting and pathological cases}
In this section, we consider pathological behaviors that appear when  integrating the TOV equations from the boundary inwards. By Thm. 1, these pathologies do not appear in systems with thermodynamically consistent EoS. We noted already that horizons can  appear if we let pressure become negative. Here, we will focus on pathologies that arise by violating the thermodynamic integrability condition, and we will explain how they can be fixed.

\subsection{Zero temperature solutions}
The EoS employed in the study of compact stars often correspond to the limit of zero temperature. The limit $T \rightarrow 0$ is taken for constant $n_a$, rather than for constant $b_a$, resulting to a functional relation between pressure and density that is not integrable.  In what follows, we will describe the thermodynamically consistent way of taking the limit $T \rightarrow 0$ in an equation of state, and describe the associated singular solutions to the TOV equations.

For concreteness, we will employ the original  Oppenheimer-Volkoff EoS \cite{OV}  that describes a
single species of free fermions with mass $m_f$. Since we consider a single species of fermions, we drop the index $a$ in $b_a$ and $N_a$. The EoS for an ideal gas of free relativistic fermions is

\begin{eqnarray}
n &=& \frac{8D}{m_f} \left[ \theta^{3/2} F_{\frac{1}{2}}(\theta, b-\theta^{-1}) +  \theta^{5/2} F_{\frac{3}{2}}(\theta, b- \theta^{-1}) \right] \label{nff}\\
P &=& \frac{16D}{3} \left[ \theta^{5/2} F_{\frac{3}{2}}(\theta, b-\theta^{-1}) + \frac{1}{2} \theta^{7/2} F_{\frac{5}{2}}(\theta, b - \theta^{-1}) \right] \label{pff} \\
\rho &=& m_f n +  8D \left[ \theta^{5/2} F_{\frac{3}{2}}(\theta, b-\theta^{-1}) +  \theta^{7/2} F_{\frac{5}{2}}(\theta, b - \theta^{-1}) \right] \label{rff},
\end{eqnarray}
where    $\theta = T/m_f$ and $D = \frac{m_f^4}{8 \pi^2 \hbar^3}$. $F_{\alpha}$ stands for the  generalized Fermi integral
\begin{eqnarray}
F_{\alpha}(\theta, s) = \int_0^{\infty} dx \frac{x^{\alpha} \sqrt{2+ \theta x}}{e^{x - s}+1}. \label{finteg}
\end{eqnarray}

The regime of highly degenerate fermions corresponds to $\theta<< 1$ and $b>> 1$.  For $\theta<<1$ and $s <0$,  $F_{\alpha}(\theta, s)$ is  suppressed exponentially:  $F_{\alpha}(\theta, s) \sim e^{-|s|}$. In Eq. (\ref{nff}---\ref{rff}), $s = b - \theta^{-1}$. We will employ the variable $Y :=  b\theta$. Then, $F_{\alpha} \sim e^{-b (Y^{-1}-1)}$ for $Y <1$.

It is convenient to view the thermodynamic variables  (\ref{nff}---\ref{rff}) as functions of $Y$ and $b$. Hence,   Eqs. (\ref{nff}---\ref{rff}) become
\begin{eqnarray}
n = \frac{D}{m_f} \nu(Y, b), \hspace{0.5cm}
P = D \sigma(Y, b), \hspace{0.5cm}
\rho = D \psi(Y, b).
\end{eqnarray}

For $b >> 1$, the functions $\nu, \sigma, \psi$
 drop sharply as  $Y $  decreases from $Y > 1$ to  $Y< 1$; the width of the transition region is of order $b^{-1}$.  In the limit $b \rightarrow \infty$, $\nu, \sigma$, and $\psi$ vanish for $Y < 1$, while for $Y> 1$ we can use the approximation

\begin{eqnarray}
F_{\alpha}(\theta,s) \simeq \int_0^{b-\theta^{-1}} x^{\alpha} \sqrt{2+ \theta x}.
\end{eqnarray}
In this limit, the functions $\nu, \sigma$  and $\psi$ depend only on $Y$,

\begin{eqnarray}
\nu(Y, b) &=& \nu_0(Y) := \frac{8}{3} (Y^2 - 1)^{3/2} \label{nfff}\\
\sigma(Y, b) &=& \sigma_0(Y) := \frac{1}{3} Y\sqrt{Y^2-1} (2Y^2-5) +  \sinh^{-1}\sqrt{Y^2-1} \\
\psi(Y, b ) &=&   \psi_0(Y) := Y\sqrt{Y^2-1} (2Y^2-1) -  \sinh^{-1}\sqrt{Y^2-1} \label{ufff}.
\end{eqnarray}

The above expressions for $n, P$ and $\rho$ are   the dominant terms in the thermodynamic variables at the limit of arbitrarily small (but non-vanishing) temperature, and arbitrarily large (but finite) $b$.   They are exact up to terms of order $b^{-1}$.


Next, we define the dimensionless variables $z = \sqrt{4\pi D}m$, and $x = \sqrt{4 \pi D} r$, and we express  Eqs. (\ref{TOV}---\ref{dN}) as
\begin{eqnarray}
\frac{dz}{dx} &=& x^2 \psi(Y, b)  \label{Yw1a}\\
\frac{dY}{dx} &=& - \frac{Y [z+x^3 \sigma(Y, b)]}{x^2 (1 - \frac{2z}{x})}. \label{Yw2a}
\end{eqnarray}
We integrate Eqs. (\ref{Yw1a}---\ref{Yw2a}) from the boundary $x = x_B := \sqrt{4 \pi D} r_B$ inwards, for constant $b >> 1$ and with initial conditions   $Y(x_B) = 1$ (zero pressure at the stellar surface) and
  $z(x_B) = z_B$, where $z_B :=   \sqrt{4\pi D}M$. The local temperature at the stellar surface is  $m_f/b$, hence,
  \begin{eqnarray}
  T_{\infty} = \frac{m_f}{b} \sqrt{1 - \frac{2z_B}{x_B}}.
\end{eqnarray}

By lemma 3, there is a point $x_2 < x_B$, such that  $dY/dx > 0$ for $x< x_2$. The solution continues to $x = 0$, where $Y = 0 $. By continuity, there is a point $x_c < x_2$, such that $Y(x_c) = 1$.
 For $b >> 1$, $\sigma(Y, b)$ and $\psi(Y, b)$  are exponentially suppressed as $x$ approaches the singularity. In fact, numerical integration stops before $x = 0$, because the values of $\sigma$ and $\psi$ reach machine precision.

Integrating Eq. (\ref{Yw1a})  from $x_c$ to $0$, we obtain
\begin{eqnarray}
z(0) - z(x_c) = \int_0^{x_c} dx x^2 \psi(Y(x), b). \label{w0w0}
\end{eqnarray}
where $z_0 = z(0)$. Since $dY/dx > 0$ for $x < x_c$, $Y \leq 1$ in $[0, x_c]$. For $Y \in[0, 1]$ and $b >> 1$, $\psi$ is an increasing function of $Y$ at constant $b$. Hence, $\psi[Y(x), b]$ in Eq. (\ref{w0w0}) is bounded above by $\psi(1, b)$. By Eq. (\ref{rff}), $\psi(1, b) = c_1 b^{-3/2} + O(b^{-5/2})$, where $c_1 = 8 \int_0^{\infty} dx\sqrt{2x}(e^x+1)^{-1} \simeq 7.67 $. We conclude that
\begin{eqnarray}
|z(0) - z(x_c)| < \frac{c_1 x_c^3}{3b^{3/2}}. \label{wxwx2}
\end{eqnarray}
As $b\rightarrow \infty$, $x_c$ becomes $b$-independent and it is determined by using the limiting expressions (\ref{nfff}---\ref{ufff}) in the TOV equation (\ref{Yw2a}). Hence, $z(x_c)$ and $z(0)$ coincide up to terms of order $b^{-3/2}$.

The total number of particles contained in the ball $x < x_c$ is
\begin{eqnarray}
N_0 =  \frac{1}{\mu \sqrt{4 \pi D}} \int_0^{x_c} dx \frac{x^2 \nu[Y(x),b]}{\sqrt{1-\frac{2w(x)}{x}}}.
\end{eqnarray}
For $x \in [0, x_c]$, $\nu(Y, b) \leq \nu(1, b)$ and $z(x) < 0$. It follows that
\begin{eqnarray}
N_0 <   \frac{c_1 x_c^3}{6\mu \sqrt{ \pi D} b^{3/2}}. \label{nbound}
\end{eqnarray}

Eqs. (\ref{wxwx2}) and (\ref{nbound}) imply that the spacetime geometry for $x < x_c$ can be approximated by a vacuum solution of Einstein's equation, namely,  Schwarzschild solution with negative mass $M_0 = -  \sqrt{4\pi D} z_0$, where $z_0 = - z(0) > 0$. In this approximation, $z(x) = - z_0$, and
\begin{eqnarray}
Y = \sqrt{\frac{1+2z_0/x_c}{1+2 z_0/x}}. \label{Yx0}
\end{eqnarray}
for all $x < x_c$. The approximation is accurate to order $b^{-3/2}$, and it becomes exact in the limit
 $b \rightarrow \infty$.
Since $L = T_{\infty} b/(Ym_f)$,  the limiting behavior of $Y(x)$ near $x = 0$ leads to the identification
\begin{eqnarray}
\eta = \frac{T_{\infty} b}{m_f(4 \pi D)^{1/4}} \sqrt{\frac{2z_0}{1+\frac{2z_0}{x_c}}}
\end{eqnarray}

\begin{figure}[tbp]
\includegraphics[height=5.5cm]{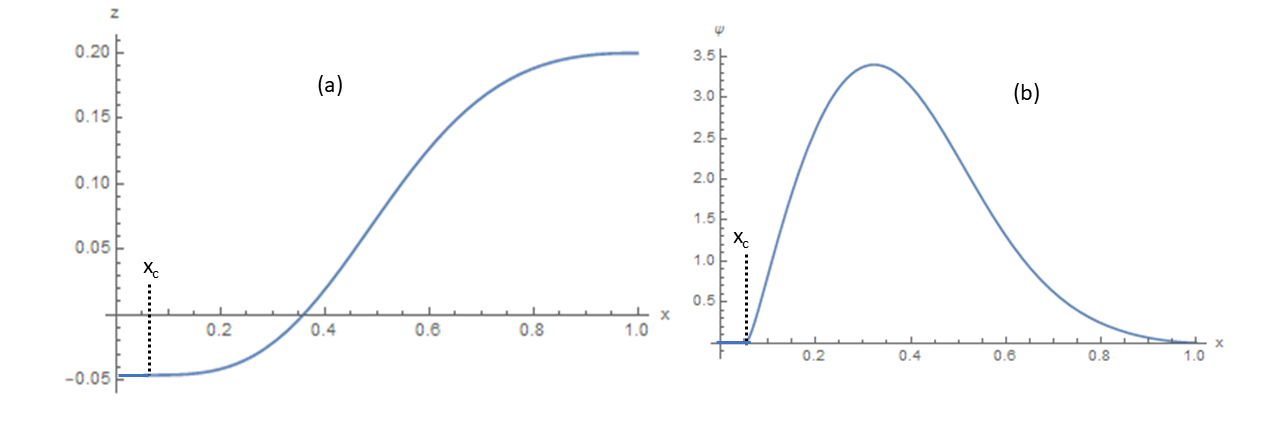} \caption{ \small  A singular solution to Eqs. (\ref{Yw1a}---\ref{Yw2a}) for $z_B = 0.15$, $x_B = 1$, at the limit $b >> 1$. (a) The dimensionless mass $z$ is plotted as a function of the dimensionless radius $x$. (b) The dimensionless energy density $\psi$ is plotted as a function of $x$.  The density vanishes for $x < x_c \simeq 0.053$. For $x < x_c$, the mass function is constant, and the geometry is that of a negative-mass Schwarzschild solution. } \label{singb}
\end{figure}

\subsection{Non integrable EoS }
Consider an EoS $P = f(\rho)$. If $f$ is a smooth function, non-integrability is due to the asymptotic behavior of $f$ at infinity or near zero. As shown in Sec. 2.4, if $f$ grows more slowly than a linear function at large $\rho$, the energy density diverges at a finite value of temperature, say, $T_{div}$. Hence, if the integration of the TOV equations leads to a value $T_{div}$ at some point $r_{div} > r_2$ (recall that $r_2$ is the point where $\partial T /\partial r$ becomes an increasing function), the energy density will diverge there.

This pathology is easy to fix, as it arises from the asymptotic behavior of the EoS as $\rho \rightarrow \infty$. We only have to modify the EoS for ultra-high temperatures, well beyond any regime that is currently known, so that condition (\ref{limeos}) applies. Then, the density divergence at high densities will be removed.

In the opposite regime, the faster-than-linear vanishing of $f$ as $\rho \rightarrow 0$ leads to a zero for $\rho$ at finite temperature, say, $T_0$. If $T_0$ appears in the integration of the TOV equations (usually at $r < r_2$), then the integration stops, as $\rho$ cannot take negative values. This is a very common behavior  of many popular EoS employed in the study of compact stars. This is more difficult to correct, because the low density regime is familiar to us, and we cannot impose arbitrary mathematical conditions.

For concreteness, we will consider a popular EoS in the study of compact stars, that originates from Gratton
 \cite{Gra64}
\begin{eqnarray}
\rho = \rho_0 \left(\frac{ P}{\rho_0}\right)^{s} + 3  P \label{gratton},
\end{eqnarray}
where $\rho_0$ is a reference pressure  and $0 < s < 1$. The EoS (\ref{gratton}) is polytropic at densities smaller than $\rho_0$; $s = \frac{n}{n+1}$ where $n$ is the usual polytropic index. For  $\rho >> \rho_0$, the EoS describes radiation.

We solve Eq. (\ref{dpt}) to obtain
\begin{eqnarray}
P &=& \rho_0 \left[\frac{ (T/T_0)^{4(1-s)} - 1}{4}\right]^{\frac{1}{1-s}} \label{PGr}\\
\rho &=& \frac{1}{4}\rho_0 \left[3 (T/T_0)^{4(1-s)}+1  \right] \left[\frac{ (T/T_0)^{4(1-s)} - 1}{4}\right]^{\frac{s}{1-s}} \label{rGr}
\end{eqnarray}
where $T_0 > 0$ is an integration constant. For $T = T_0$, $\rho = P = 0$.
Hence, for Gratton's EoS leads the integration of the TOV equations towards the center stops at a value of $r$ where the temperature $T_0$ is reached.

To correct this behavior we need to extrapolate Eqs. (\ref{PGr}) and (\ref{rGr}) to $T < T_0$.   Gratton's EoS is supposed to describe fermionic degenerate matter, so the limit of low densities (and temperature) essentially describes a cold  dilute gas of fermions of mass $m_f$. The latter is given by Eqs. (\ref{ncold}---\ref{rcold}). Hence, we need to connect Eqs. (\ref{PGr}) and (\ref{rGr}), with the EoS for a dilute gas
\begin{eqnarray}
P = C e^{-m_f/T} T^{5/2}, \hspace{2cm} \rho =   C e^{-m_f/T}T^{3/2} (m_f +\frac{3}{2}T),
\end{eqnarray}
for some constant $C>0$. Since $\rho = T(\partial P/\partial T)  - P$, it suffices to connect the functions $P(T)$ at some point $T_1 > T_0$. Since we have two parameters $C$ and $T_1$, we can always choose $P$ and its first derivative to be continuous at $T_1$.


In  the EoS for a cold dilute gas of fermions, pressure and energy density drop exponentially as $e^{-m_f/T}$ as temperature decreases. In this sense, the condition $T = T_0$ is analogous to   the condition $Y = 1$ of the model in Sec. 5.1. Hence, we can treat the associated solutions to the TOV equation the same way. Let $r_c < r_2$ be a point such that $T(r_c) = T_0$ in a solution. The total mass contained in the ball $r \in [0, r_c)$ is suppressed by a factor  $e^{-m_f/T_0}$. If $m_f >> T_0$, it is a good approximation to treat the solution in $[0, r_c)$ as a vacuum solution, with negative mass given by $m(r_c) < 0$. Hence, we will obtain a solution similar to that of Fig.   (\ref{singb}).

\section{Dynamical stability}

The vanishing of the pressure at the center of a singular solution to the TOV equation may appear counter intuitive, when compared with the usual insight that compact stars are stabilized as a result of high central pressure. Indeed, one would expect that all solutions with positive pressure gradients  are dynamically unstable. However, this physical intuition follows from the study of regular solutions. Singular solutions have the novel feature that the singularity repulses the interior matter layers. The issue is whether this repulsion suffices to compensate for the lack of central pressure, as fa stability is concerned. In this section, we undertake a preliminary investigation of this issue, by studying adiabatic radial perturbations.

\subsection{Adiabatic radial perturbations}
 A stationary solution to  Einstein's equations is (linearly) dynamically stable, if no linearized perturbation admits runaway solutions. Here, we discuss radial perturbations that are typically characterized by the strongest instabilities.

Consider spherically symmetric metrics of the form
\begin{eqnarray}
ds^2 = - L^2(r,t)dt^2 + \frac{dr^2}{1-\frac{2m(r, t)}{r}} + r^2 (d\theta^2 + \sin^2\theta d \phi^2),
\end{eqnarray}
with time-dependent lapse $L(r, t)$ and mass function $m(r, t)$ that perturb a static solution to Einstein's equations.

The linearized perturbations to Einstein's equation can be expressed in terms of an appropriately chosen function $f(, t)$ that satisfies a hyperbolic equation \cite{MTW, Yabu}
\begin{eqnarray}
W\ddot{f} = (Sf')' +Qf \label{pert1}
\end{eqnarray}
where $S, Q$ and $W$ are functions of $\xi = - \log(r/r_B)$ that are determined by the equilibrium solution. A dot denotes differentiation with respect to $t$ and a prime denotes differentiation with respect to $\xi$.

For adiabatic perturbations, the functions $W, S$ and $Q$ are given by  \cite{MTW}
\begin{eqnarray}
W(\xi) &=& \frac{1}{4\pi r_B^3}e^{3\xi} \frac{L(v+w)}{(1-u)^{3/2}} > 0,\\
S(\xi) &=&  \frac{1}{4\pi r_B^5} e^{5\xi}  \frac{L^3 \Gamma_1 w}{ \sqrt{1-u}} >0,  \\
Q(\xi) &=& \frac{1}{4\pi r_B^5}e^{5\xi}  \frac{L^{3}(v+w)}{\sqrt{1 - u}}\left[ (t')^2 +4t'  - \frac{2w}{1-u}\right]. \label{Qxi}
\end{eqnarray}
where $L, u, v, $  $w$ refer to the equilibrium solution; $\Gamma_1 := (\partial \log P/\partial \log n)_{\frac{s}{n}} $ is the fluid's adiabatic index.

For oscillatory perturbations  $f(r) = f_{\omega}(r)e^{-i \omega t}$ with frequency $\omega$, Eq. (\ref{pert1}) becomes
\begin{eqnarray}
 (Sf_{\omega}')' +Qf_{\omega} +\omega^2 W f_{\omega}= 0.\label{pert2}
\end{eqnarray}
Eq. (\ref{pert2}) is a Sturm-Liouville equation. The usual boundary conditions for $f_{\omega}$ are $[Sf'_{\omega}](0) = 0$, and $ f_{\omega}(\infty) = 0$ \cite{MTW}. For these boundary conditions, the Sturm-Liouville operator $L$, defined by $Lf :=  \frac{1}{W} (Sf')' +Qf $ is self-adjoint. Hence, its eigenvalue $\omega^2$ are real valued.

Suppose we order the eigenvalues of $L$  as $\omega_0^2 < \omega_1^2 < \ldots < \omega_n^2 < \ldots \rightarrow \infty$.   A negative eigenvalue $\omega^2$ signifies a mode growing unboundedly; hence, dynamical instability.  It follows that   $\omega_0^2 > 0$ is  a necessary and sufficient condition for dynamical stability.

The smallest eigenvalue $\omega_0$  can be determined from a variational principle,
\begin{eqnarray}
\omega_0^2 = \min_{f \in {\cal K}} R[f] \label{minimmm}
\end{eqnarray}
where $R[f]$ is the Rayleigh-Ritz functional
\begin{eqnarray}
R[f] :=\frac{\int_0^{\infty}d\xi (Sf'^2 - Qf^2)}{\int_0^{\infty}d\xi Wf^2}.
\end{eqnarray}
The minimum in Eq. (\ref{minimmm}) is taken over the set ${\cal K}$ of  {\em all}   differentiable function $\zeta$, with finite $R[f]$, that satisfy  the boundary conditions.
Hence, dynamical stability requires that
\begin{eqnarray}
\int_0^{r_B}dr (S\zeta'^2 - Q\zeta^2) > 0, \label{condsta}
\end{eqnarray}
for {\em any} $\zeta \in {\cal K}$ .

 It is well known that in regular solutions, instability appears at high central temperatures and pressures. This can be seen from the ratio $Q/W$ as $\xi \rightarrow \infty$,
\begin{eqnarray}
\frac{Q}{W} \rightarrow 8 \pi \bar{L} \bar{P},
\end{eqnarray}
where we use the overbar to denote the value of a variable at the center.

This ratio gives the strength of the negative contribution to $R[f]$ near the center. In contrast, the ratio $S/W$ as $\xi \rightarrow \infty$ is proportional to $\bar{P}/(\bar{\rho}  + \bar{P})$, and it is bounded with increasing central temperature. This suggests that for solutions with sufficiently large central temperature, the negative contribution from $Q$ dominates, and negative eigenvalues $\omega_0^2$ occur.

\subsection{Enhanced stability}

  The behavior of the Rayleigh-Ritz functional is very different for singular solutions. First, we note that for solutions to the TOV equations, $t' < 0$ implies that $t' > -4$. Indeed, the latter inequality is equivalent to $w > -4 + \frac{7}{2}u$, which is always valid, because $u<0$ for $t' < 0$. It follows that $Q(\xi) < 0$ for $\xi> \xi_2$.

The function $Q$ remains negative up to a point $\xi_s \in (\xi_1, \xi_2)$. The point $\xi_s$ can be characterized by  $\delta(\xi_s)$, where  $\delta := |u|/w$ on it; $\delta$ varies between 0 at $\xi_1$ and 2 at $\xi_2$. Substituting Eq. (\ref{TOV3}) into Eq. (\ref{Qxi}), we find that $Q$ has the same sign with the quantity
\begin{eqnarray}
-\frac{7}{4}u^2 + 2u+w^2 +w - uw = (-\frac{7}{4}\delta^2 + 2\delta +1) w^2+ 2 (1-\delta) w.
\end{eqnarray}
This quantity is always positive if $(1-\delta) > 0$ and $ -\frac{7}{4}\delta^2 + 2\delta +1 > 0$.  It  is always negative if $(1-\delta) <  0$ and $ -\frac{7}{4}\delta^2 + 2\delta +1 < 0$. Combining these inequalities, we find that  $\delta (\xi_s)$   lies  between      $1$ and $\frac{1}{7}(4+\sqrt{11}) \simeq 1.52$.  It follows that $Q$ remains negative in a large neighborhood of the maximum-temperature point $\xi_2$.

We conclude that the interior layers of the singular solution ($\xi > \xi_s$) always contribute a positive term to the Rayleigh-Ritz functional. Hence, the repulsive singularity at the center {\em enhances dynamical stability}.
Dynamical instability necessitates  a negative contribution to the Rayleigh-Ritz functional from $\xi < \xi_s$ that overcome the positive contributions from the inner layers. This is   possible if the maximum value of $u$ is very close to unity, as $Q$ grows with $(1 - u)^{-5/2}$ and $S$ with $(1-u)^{-1/2}$. Hence, instabilities are correlated with the existence of surfaces of high blue-shift in the outer layers. Since many singular solutions do not have such surfaces, we expect that stable singular solutions are generic.

\subsection{An example}

As an example, we consider the stability of the solutions that were studied in Sec. 5.1.  At the limit of very large $b$, the $b$-dependence factors out, so that the space of solutions $\Gamma$ is two-dimensional. We parameterize this space by the dimensionless mass $w_B \in[0, \infty)$ and the compactness $u_B \in [0,1 )$. Regular solution define an one-dimensional submanifold of $\Gamma$ that is plotted in Fig. 5.

The OV limit for this model corresponds to $z_B = z_{OV}\simeq 0.153$. For $z_B > z_{OV}$, there are no regular solutions. For $z_B < z_{OV}$, there are regular solutions. For $z_B < z_1 \simeq 0.08$, there is only one regular solution for each $z_B$, for $z_B \in (z_1, z_{OV})$ there are more than one solutions.

\begin{figure}[tbp]
\includegraphics[height=5.5cm]{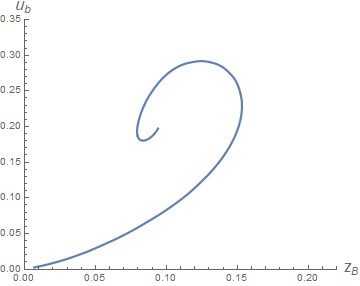} \caption{ \small  The curve of regular solutions for the model of Sec. 5.1. }
\end{figure}

We constructed the Reyleigh-Ritz functional for this class of solutions. For each $(z_B, u_B)$, we estimated the smallest eigenvalue $\omega_0^2(z_B, u_B)$ using a variational method. We found it convenient to employ the function
\begin{eqnarray}
F(z_B, u_B) = \exp\left[\frac{\omega_0^2(z_B, u_B)}{4\pi D^2}\right] - 1.
\end{eqnarray}
If $F(z_B, u_B) > 0$, then the solution with $(z_B, u_B)$ is stable, otherwise, it is unstable.
In Fig. 6, we plot $F(z_B, u_B)$ as a function of $u_B$ for  representative values of $z_B$.

\begin{figure}[tbp]
\includegraphics[height=4.5cm]{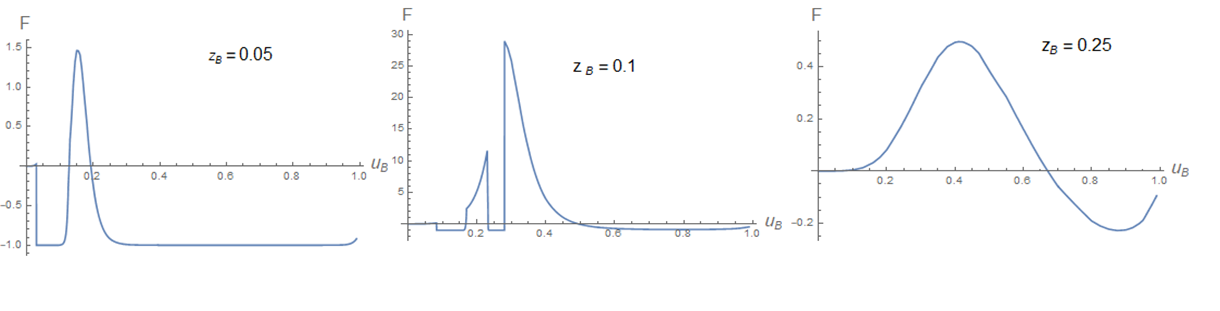} \caption{ $F(z_B, u_B)$ as a function of $u_B$ for different values of $z_B$. }
\end{figure}
Our conclusions from the numerical study of $F(z_B, u_B)$ are the following.
\begin{enumerate}
\item For all $z_B$, the solutions are stable for sufficiently small $u_B$, and unstable for sufficiently large $u_B$.
\item For $z_B < z_{OV}$, there is one or more islands of stability between regions of $u_B$ that describe unstable solutions.
\item For $z_B < z_{OV}$, $F$ is discontinuous at points $u_B$ that correspond to regular solutions.
\item For $z_B > z_{OV}$, $F$ is continuous, and there is a single point of transition from stable to unstable solutions.
\end{enumerate}

\subsection{Thermodynamical stability}
A solution to the TOV equation is physically meaningful  if it is both dynamically and thermodynamically relevant. We showed that dynamical stability is  plausible for singular solutions, even though the analysis of more general types of perturbation are needed. Thermodynamic stability is a more complex issue, in fact, the thermodynamic consistency of solutions to the TOV equation is one of the main motivations for this work \cite{AnSav12}.

We will undertake the thermodynamic analysis in a future work. The reason is that such an analysis requires the introduction of entropy associated to the singularity \cite{AnSav12}. The arguments in support for this singularity entropy goes beyond the scope of this work, and it does involve a degree of conjecture, as it goes beyond classical General Relativity. We only note that such an analysis was undertaken in Ref. \cite{AnSav12} for self-gravitating radiation. There, it was shown that singular solutions are not thermodynamically stable for masses $M$ smaller than the Oppenheimer-Volkoff limit $M_{OV}$, and that some thermodynamically stable singular solutions exist for $M > M_{OV}$. We believe that this result can be generalized to all solutions of the TOV equation that were considered in this paper.

  \section{Conclusions}
 Our results  provide a complete characterization of {\em all}  solutions to the TOV equation, including singular ones. All singular solutions share the same structure and have a curvature singularity at the center. The singularity strongly repulses any matter that approaches it. The repulsion enhances stability, and also results to only a subset of measure zero of all geodesics reaching the singularity.

We will explore the possibility that singular solutions to the TOV equations correspond to actual astrophysical objects in a different publication. Here, we summarize the reasons why these solutions are of  significant interest. They are generic solutions a paradigmatic equation of relativistic astrophysics, and for this reason, it is natural to expect that they appear in final stages of gravitational collapse. They describe a geometry in which  a region with matter interpolates between
 a negative-mass Schwarzschild spacetime at the center to a positive-mass Schwarzschild spacetime at infinity.
They provide an tractable model of a spacetime with timelike singularity that is bounded acceleration complete.  Their stability analysis is intriguingly  novel: stability is enhanced by the repulsive singularity at the center and it is disrupted by high blue-shift surfaces at the outer layers of the solution.

\end{document}